    \documentclass[runningheads]{llncs}
    \usepackage[T1]{fontenc}
    \usepackage{comment}
    \usepackage{graphicx}
    \usepackage{minted}
    \usepackage{multirow}
    \usepackage{amsmath}
    \usepackage{comment}
    \usepackage{amssymb}
    
    \usepackage[normalem]{ulem}  
    \usepackage{xcolor}          

\begin{document}
    \title{Improving Router Security using BERT}
    
    \titlerunning{Improving Router Security using BERT}
    %
    \author{John Carter\inst{1} \and
    Spiros Mancoridis\inst{1} \and
    Pavlos Protopapas\inst{2} \and
    Brian Mitchell\inst{1} \and
    Benji Lilley\inst{1}
    }
    \authorrunning{Carter et al.}
    %
    \institute{Drexel University, Philadelphia, PA, USA
    \email{\{jmc683,mancors,bmitchell\}@drexel.edu} \and
    Harvard University, Cambridge, MA, USA\\
    \email{pavlos@seas.harvard.edu}}
    \maketitle              
    \begin{abstract}
Previous work on home router security has shown that using system calls to train a transformer-based language model built on a BERT-style encoder using contrastive learning is effective in detecting several types of malware, but the performance remains limited at low false positive rates. In this work, we demonstrate that using a high-fidelity eBPF-based system call sensor, together with contrastive augmented learning (which introduces controlled mutations of negative samples), improves detection performance at a low false positive rate. In addition, we introduce a network packet abstraction language that enables the creation of a pipeline similar to network packet data, and we show that network behavior provides complementary detection signals—yielding improved performance for network-focused malware at low false positive rates. Lastly, we implement these methods in an online router anomaly detection framework to validate the approach in an Internet of Things (IoT) deployment environment.
    
    \end{abstract}
    \section{Introduction}
    
    \indent Home routers serve as an important component in any home network as they are the gateway to the outside world. Nevertheless, home network security is often overlooked by security analysts despite the rise of connected and unmonitored Internet of Things (IoT) devices. Related work by Carter et al. \cite{carter2025contrastbert} demonstrates that the use of a specialized language called \textbf{sys2vec} for the embedding of calls to the Linux kernel-level operating system (OS) is effective for protecting home routers by using unsupervised anomaly-detection. However, the approach described in that work showed limited robustness at low false positive rates, which is a critical operating regime for always-on router deployments.
    
    \indent This work replicates the environment and the malware in that research to illustrate three extensions to that approach. The first is the introduction of \emph{contrastive augmented} learning to the BERT-style language-model training process, where controlled mutation of negative samples improves robustness at low false positive rates compared to using contrastive learning alone. The second is the introduction of a simple network packet abstraction language that enables us to apply a similar pipeline to that of the system calls and capture complementary network-level behaviors, which leads to improved detection for network-focused malware at low false positive rates. We find that for the best detection, both a system call detector and a network packet detector should be deployed concurrently. The last contribution is an online anomaly detection framework using the methodology described in this work, which illustrates how the approach can operate in a practical deployment.
 
    \indent The system calls and network packets are collected during periods of strictly benign behavior, as well as separately during periods of malware infection. The packet data include all traffic on its network. In order to isolate malicious behavior, rather than using a mixture of benign and malicious behavior, only system calls and packets with process identifiers (PIDs) matching those generated by the malware during their execution are considered during evaluation. For deployment, where PID filtering is not available, an exponential moving average (EMA) is used to smooth the continuous stream of observations, allowing the detector to operate without requiring PID filtering while reducing false positives.
    
    \indent The system call data are transformed using sys2vec, as described in the previous work by Carter et al. \cite{carter2025contrastbert}. In contrast, the packet data are first generalized using a simple packet abstraction language. To generalize, we select a portion of the available packet header fields, such as specific IP addresses and port numbers, and put them into abstract categories before concatenating them together. Following the abstraction procedure, each distinct packet abstraction (word) seen in both benign and malware data is embedded into a 64-dimensional space using net2vec, which provides the same form of embeddings as sys2vec but is used for our packet language. To avoid out-of-vocabulary issues, the vocabulary is built from all packet abstractions observed across the full dataset, without using any malware-specific labels during training.
    
\begin{figure}[!t]
    \centering
    \includegraphics[width=0.65\textwidth]{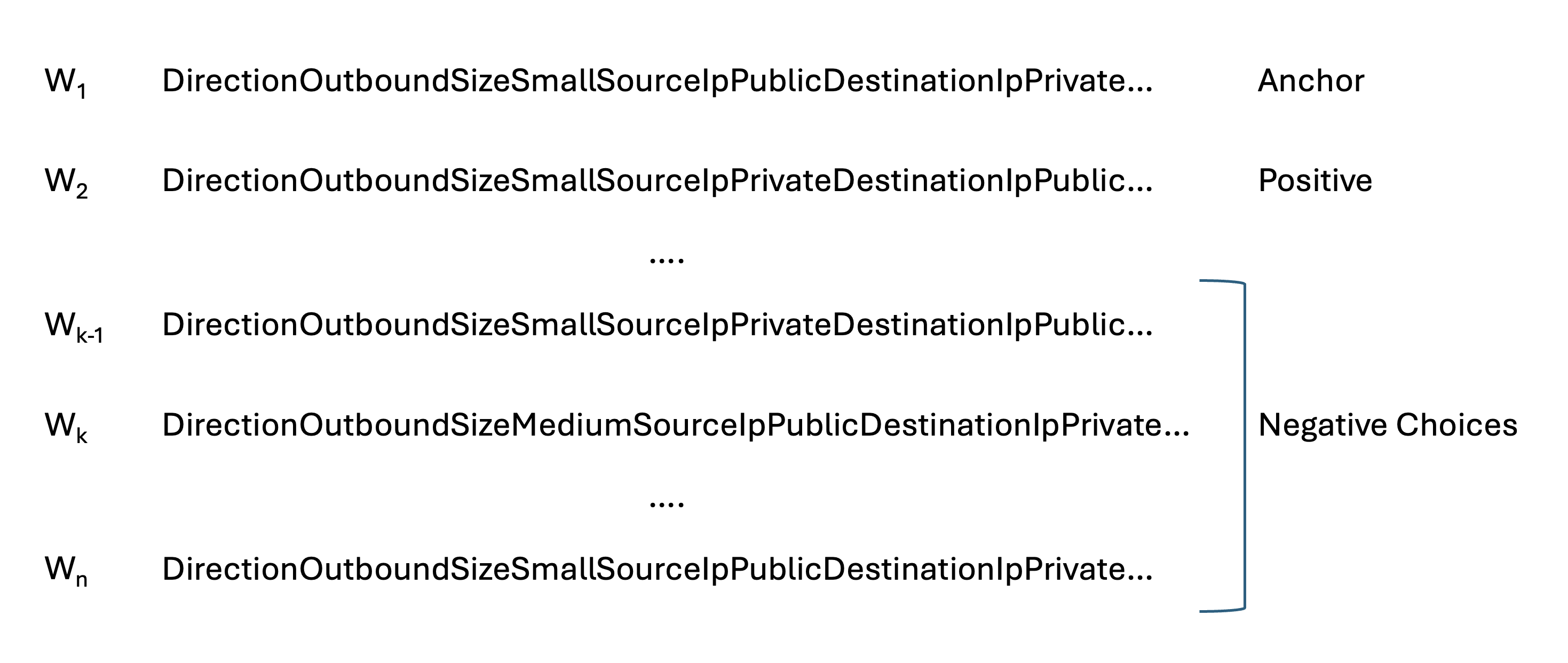}
    \caption{Example of how the negative window is selected from future windows relative to the current anchor and positive sample.}
    \label{fig:randomnegative}
    \end{figure}

    \indent The packet data are split into 100-word sentences, with a window length empirically selected to balance temporal context and vocabulary coverage.
    Anchor samples, positive samples, and negative samples were created from these sentences to feed into our contrastive augmented BERT-style model framework. 
    The choice of these samples is crucial for model training, as it attempts to bring the positive sample closer while simultaneously pushing the negative sample further away from the anchor during the learning process \cite{Schroff_2015}. 
    The triplet selection algorithm is shown in more detail in Figure \ref{fig:randomnegative} and is explained in more detail below. 
    The randomly chosen negative sample is mutated prior to being passed into the model by replacing a subset of the 100 packets in the sentence with randomly selected packets observed in the training dataset. 
    We also evaluated contrastive learning without mutation, as in \cite{carter2025contrastbert}, but found that adding controlled perturbations to negative examples improved training robustness, as summarized in Tables  \ref{tab:syscall-rates} and \ref{tab:network-rates}.
    
    \indent The paper is organized as follows. First, it describes related work and the network ecosystem that the home router serves. 
Then it describes how we developed the malware patterns according to the specifications described by Carter et al. \cite{carter2025contrastbert}. 
    Next, it examines in detail how we collect the network packets and transform them to be used by our language-model framework, and summarizes how the system calls are transformed since this was explained in detail in prior work. 
    Afterward, it describes how both contrastive augmented language models are trained and evaluated, which is a very similar process for both system calls and network packets. 
    Next, the online anomaly detection framework is described, which provides a deployable version of the research presented. 
    Lastly, it identifies avenues for future work and summarizes our findings.
    
    \section{Related Work}
    \subsection{Behavioral Malware Detection}
    Behavioral malware detection emerged several years ago as a response to the rise of zero-day and obfuscated malware, which has made traditional signature matching anti-virus protection less effective \cite{10757249}. The reason for this is because antivirus detection relies on having seen malware that matches the binary signatures found in its database. However, malware obfuscation techniques work to change and/or encrypt their code with each replication, like polymorphic and metamorphic malware \cite{NATSOS2025104250}.
    
    Behavioral malware detection seeks to model the benign behavior of a device in order to flag behavior deviating from this benign model as malicious, meaning that training requires only benign data while malware traces are used solely for evaluation. The benefit of this approach is that no prior knowledge of malware is necessary and only benign data are necessary for model training, which is often readily available, in contrast to realistic and usable malware data. Several shallow machine learning models have been used for anomaly detection, such as one-class SVMs, autoencoders, and random forests \cite{10.1145/2500853.2500857} \cite{9439459}.  More recently, sequence-based approaches have been explored to capture temporal structure in device behavior, motivating the use of transformer-style models in subsequent sections of this work.
    
    Although there has been work in the area of online anomaly detection, the area has been much less explored than traditional offline anomaly detection using static datasets.  There exists an online anomaly detection study that explores the effectiveness of a framework for testing the quality of data streamed in a large telecommunication system \cite{Rettig2019}.  This is relevant to our work in terms of the use of data streams, but differs in its application as it does not pertain to malware detection. Similarly, another study examines online anomaly detection for sensor systems and highlights their requirements such as accuracy, robustness, resource efficiency, and performance \cite{YAO20101059}.  These studies help frame the broader landscape of online detection, but they do not address the specific challenges posed by router-level behavioral monitoring, which motivates the online component of our work.

    \subsection{Large Language Models for Malware Detection}
    The popularity of large language models (LLMs) has grown significantly as commercial chatbot applications are used for tasks such as writing and coding. 
    Although there has been an extensive amount of research into training these models using natural spoken languages, there has been comparatively less exploration of applying LLM-style sequence modeling to specialized machine-generated languages such as system calls or packet abstractions.
    
    \indent Significant research exists to harness the power of LLMs to increase results in a variety of areas, such as ransomware detection \cite{Zhou_Liu_Meng_Tao_Tian_Yao_Li_Han_Chen_Yang_2025} and IoT malware detection using information from network packets \cite{10757249}. 
    These approaches differ from ours in that we analyze packet streams collected directly from a home router ecosystem and generate embeddings using a task-specific packet abstraction language, rather than relying on generic LLM tokenization or pre-trained embedding spaces.
    There have also been efforts to detect Android malware using LLMs, which attempts to model semantic dependencies within Android application packages (APKs) \cite{10979936}.
    
    \indent Lastly, there has been recent work on the detection of malware on Linux devices using representative call systems from the device and a range of LLMs, including BERT, GPT-2, and Mistral \cite{10773857}. 
    This body of work overlaps with our studies, although it has some key differences.  
    One is that their work uses pre-trained models with a classification layer on top \cite{10773857} for model fine-tuning, whereas our models are trained using run-time system call or network packet data. 
    Additionally, their models primarily perform supervised binary classification, whereas our approach is unsupervised anomaly detection, which is better aligned with detecting zero-day or stealthy attacks.
    
    \subsection{Contrastive Learning}
    Contrastive learning has been applied to several domains, such as anomaly detection in graphs \cite{luo2022deep} and in images \cite{10325644}.
    It has also been used recently in conjunction with LLMs for medical research \cite{WU2025127241}, code authorship \cite{10.1145/3705300}, and training LLMs to respond in a certain way to align with the intent of a user \cite{Gao_Das_2024}. 
    These are just a few examples of widely divergent fields that are all harnessing the power of transformer-based models using contrastive learning. 
    Similarly, contrastive methods with augmented or perturbed negatives have been used in a wide array of problem domains, such as facial recognition \cite{Kim_Song_2021} and natural language processing \cite{miao2021simplecontrastiverepresentationadversarial}, although to our knowledge, this type of learning has not yet been applied to the detection of behavioral anomalies in router-level system call or packet data.
    
    \indent There are many variants of contrastive loss, such as mean-shifted contrastive loss, introduced by Reiss et al. \cite{Reiss_Hoshen_2023}, as well as triplet loss \cite{Schroff_2015}. 
    Triplet loss is a popular loss function for machine learning models in a wide range of model applications, such as computer vision \cite{Schroff_2015}, and aims to learn meaningful data representations by comparing the distances of the three triplet vectors \cite{hoffer2018deepmetriclearningusing}. 
    Our triplet loss framework is similar to the one used by Schroff et al., in which the authors used a triplet mining strategy to align matching and non-matching faces \cite{Schroff_2015}. 
    In our setting, the same structure is used to learn representations of benign router behavior, where negatives are drawn from other benign sequences and lightly perturbed to create harder contrastive examples for anomaly detection.

    
     \section{Methodology}\label{approach}
    This section describes the methodological components of our approach, including system call collection, packet abstraction, embedding, triplet construction, contrastive augmented learning, and online anomaly detection. The experimental environment used to generate the system call and packet traces is presented later in Section~\ref{sec:ecosystem}.
    
    \subsection{System Call Collection}

System call traces are collected using a custom eBPF~\cite{nakryiko2020bpf} sensor tailored to our machine learning pipeline for malware detection. Alternative userspace tools like \texttt{ftrace} excel as general-purpose utilities for system administrators performing broad performance analysis and debugging~\cite{volpert2025ebpf}. Our sensor attaches selectively to syscall tracepoints or kprobes and extracts only the arguments, PIDs, and metadata needed for tokenization and embedding. It avoids the broader event sets such as scheduler and interrupts that \texttt{ftrace} often enables by default~\cite{findlay2020bpfsec}.

This targeted design minimizes data volume from the start. Our eBPF programs apply in-kernel filtering and aggregation before events reach userspace ring buffers for processing. This differs from \texttt{ftrace} sessions that may require post-collection processing even when filtered~\cite{cassagnes2020ebpf}. For our AI platform, this efficiency supports high-fidelity capture of syscall sequences under load, where \texttt{ftrace}'s per-CPU buffers, designed to protect system stability by throttling writes, can drop events during bursts and explicitly report these losses via \texttt{trace} and \texttt{dropped} counters if not finely tuned for our specific use case~\cite{volpert2025ebpf}.

eBPF's programmable nature further aligns with our needs. It streams filtered events via low-overhead ring buffers to userspace and preserves timing and ordering critical for behavioral malware anomaly detection~\cite{findlay2020bpfsec}. Each system call is mapped to a token using the sys2vec language introduced by Carter et al.~\cite{carter2025contrastbert}, which groups calls into semantic categories (e.g., file access, networking, process control). The resulting tokenized sequences form the input to the embedding and contrastive learning pipeline described in the following subsections. Adapting \texttt{ftrace} for ML-specific tokenization like sys2vec categories would demand extra scripting and lacks eBPF's native integration for custom schema output~\cite{cassagnes2020ebpf}.

System calls are recorded during both benign execution and controlled malware runs. For offline evaluation, we retain only calls whose PIDs correspond to malware processes so that anomaly scores reflect only malware behavior.  In deployment scenarios, where PID filtering is not available, we introduce an exponential moving average (EMA) to smooth anomaly scores over time. This architecture leverages eBPF's strengths for our specialized pipeline and delivers cleaner, lower-overhead data than repurposing a general tracer like \texttt{ftrace}~\cite{volpert2025ebpf,findlay2020bpfsec}.

    \subsection{Network Packet Collection}
    
    Network packets are similarly collected using eBPF programs attached at both the socket layer, where attribution to userspace processes is preserved, and the traffic-control (TC) layer for low-level packet inspection of all system-transiting traffic. This dual-instrumentation approach captures process-specific network behavior alongside ambient router traffic, complementing our syscall traces with complete communication context.
    
    Network packets are transformed into a discrete ``packet language’’ using a light\-weight abstraction scheme. Instead of embedding raw headers, we extract selected fields (e.g., IP addresses, port ranges, direction, and protocol identifiers) and map them into categorical groups such as \texttt{SourcePortWellKnown}, \texttt{DirectionInbound}, or \texttt{ProtoTCP}. These categorical components are concatenated to form a single token representing each packet.
    
    Network traffic provides a complementary behavioral signal that system calls alone cannot fully capture. Although network activity is ultimately generated through system calls such as \texttt{socket}, the syscall stream offers only a noisy and fragmented view of higher-level communication patterns. Key characteristics of malware—such as destination address regularity, port-selection strategies, burstiness of outbound flows, and repeated connection attempts—are directly observable at the packet level but are difficult to infer reliably from low-level syscall sequences. For this reason, packet abstractions provide a more stable and discriminative representation of network-intensive malware, especially at low false-positive operating points where syscall-only models struggle.
    
    This abstraction reduces noise from high-entropy header fields, improves generalization across devices, and enables packet traces to be treated as sequences analogous to system calls.
    Each unique packet abstraction token is then embedded using net2vec, a Word2Vec-style embedding model tailored to the packet vocabulary. The resulting embeddings provide continuous vector representations suitable for transformer-based sequence modeling.
    
    Like many types of device communication, packet traffic can be reduced to a small set of vocabulary words, each with a distinct meaning. Since all packets have a similar structure, the key to differentiating them by their function is the effective abstraction of the packet fields into meaningful categories. For example, IP addresses are often ephemeral and therefore not helpful for training a generalizable model. Similarly, granular values such as the exact byte size of a packet create a sparse vocabulary that is not conducive to model training.
    
    After preprocessing, only the protocol, source IP, source port, destination IP, destination port, size in bytes, and packet direction fields are left, which are then bucketed into more general groups. The IP addresses are abstracted into either Private, Loopback, LinkLocal, Multicast, Documentation, or default to ``Public'' if no matches are found.
    Ports are abstracted into three groups: ``WellKnown'' (0–1023), ``Registered'' (1024–49151), and ``Dynamic'' (49152–65535). Size is abstracted into Small, Medium, and Large, and direction into Inbound and Outbound. Lastly, the three protocols we use are TCP, UDP, and ICMP.

    
    From these abstractions, we train our net2vec model using Word2Vec in Gensim~\cite{rehurek2011gensim}. The final vocabulary size is 70, corresponding to 70 distinct packet-abstraction tokens observed in the collected data. Although small compared to natural languages, this vocabulary is sufficient for the number of packet-behavior patterns present in the router environment.
    
    As part of our ablation study, we also evaluated fixed random embeddings, but found that net2vec combined with a 0.1 mutation rate consistently produced the most stable detection performance across contamination rates. The mutation strategy is described further below.
    
    \subsection{Triplet Construction}
    
    System call sequences are segmented into fixed-length windows of 500 tokens, 
    while packet sequences are segmented into windows of length 100. These window 
    sizes were chosen empirically to provide sufficient temporal and contextual 
    coverage without inflating sequence length unnecessarily.
    
    For all offline experiments, malware data contain only system calls or network packets associated with malicious PIDs. This ensures that only malware-generated data are used in the evaluation, as mixing of benign and malware data yields imprecise results and the goal is to measure how far a malware process deviates from 
    the learned benign profile. It does not assume that PID information is available 
    during deployment (this is addressed separately by our online detector).
    
    From these sequences we construct triplets 
    $\{\text{anchor}, \text{positive}, \text{negative}\}$ drawn from benign data only. 
    The anchor is a benign window, the positive is the next sequential benign window, 
    and the negative is a randomly selected benign window from a later point in the 
    same PID stream. This structure encourages the model to learn stable 
    representations of benign behavior by bringing temporally adjacent windows 
    closer together while pushing unrelated windows further apart.
    
    \subsection{Contrastive Augmented Learning}
    
    To increase the difficulty of the contrastive task, we apply a lightweight mutation to each negative sample: a small fraction of its tokens is replaced with randomly selected tokens drawn from the overall training vocabulary. This is not intended as an adversarial attack, but rather as a controlled augmentation that produces harder negative examples.
    
    The motivation for mutating negative samples is to generate negatives that remain close to the benign manifold while differing in semantically meaningful ways. Standard contrastive learning draws negative windows from unrelated parts of the benign trace, but we find that some mutation to the randomly-selected negative sentence produces more stable embeddings and improves anomaly detection performance at low false positive rates relative to standard contrastive learning without mutation.
    
    \subsection{calBERT Model Architecture}
    We train two BERT-based models from scratch: one for system calls using sys2vec embeddings and one for network packets using our net2vec embeddings. The architectures are parallel so that differences in performance arise only from the underlying data types rather than differences in model capacity or training procedure. Only benign data are used for training, consistent with the anomaly-detection objective.
    
    For the system call model, we follow the contrastive augmented learning framework described above. For the SOTA' replication, we adopt a hard-negative mining strategy: for each anchor--positive pair, we sample 50 candidate windows and select the candidate with maximum embedding distance as the negative sample. 
    
    For mutation rates $m \in \{0.1, 0.2\}$, we select a future window as the negative sample and mutate $m$ proportion of its tokens by replacing them with randomly selected system calls. This produces harder negative examples and allows us to evaluate the impact of controlled augmentation on model performance.
    
    The network-based BERT model is trained similarly, using the packet abstraction language and the net2vec embeddings. For each anchor window, the subsequent window is used as the positive, and the negative is chosen as a random future window. The negative window is then mutated by replacing a fraction of the packet tokens with other packet abstractions observed in the training data. We sweep mutation rates of 0, 0.1, and 0.2 to compare their effects, where the 0\% mutation is simply a randomly-selected negative sample without any mutation. In practice, 10\% mutation provides the most stable performance.
    
    \begin{figure}[!t]
    \centering
    \includegraphics[width=0.5\textwidth]{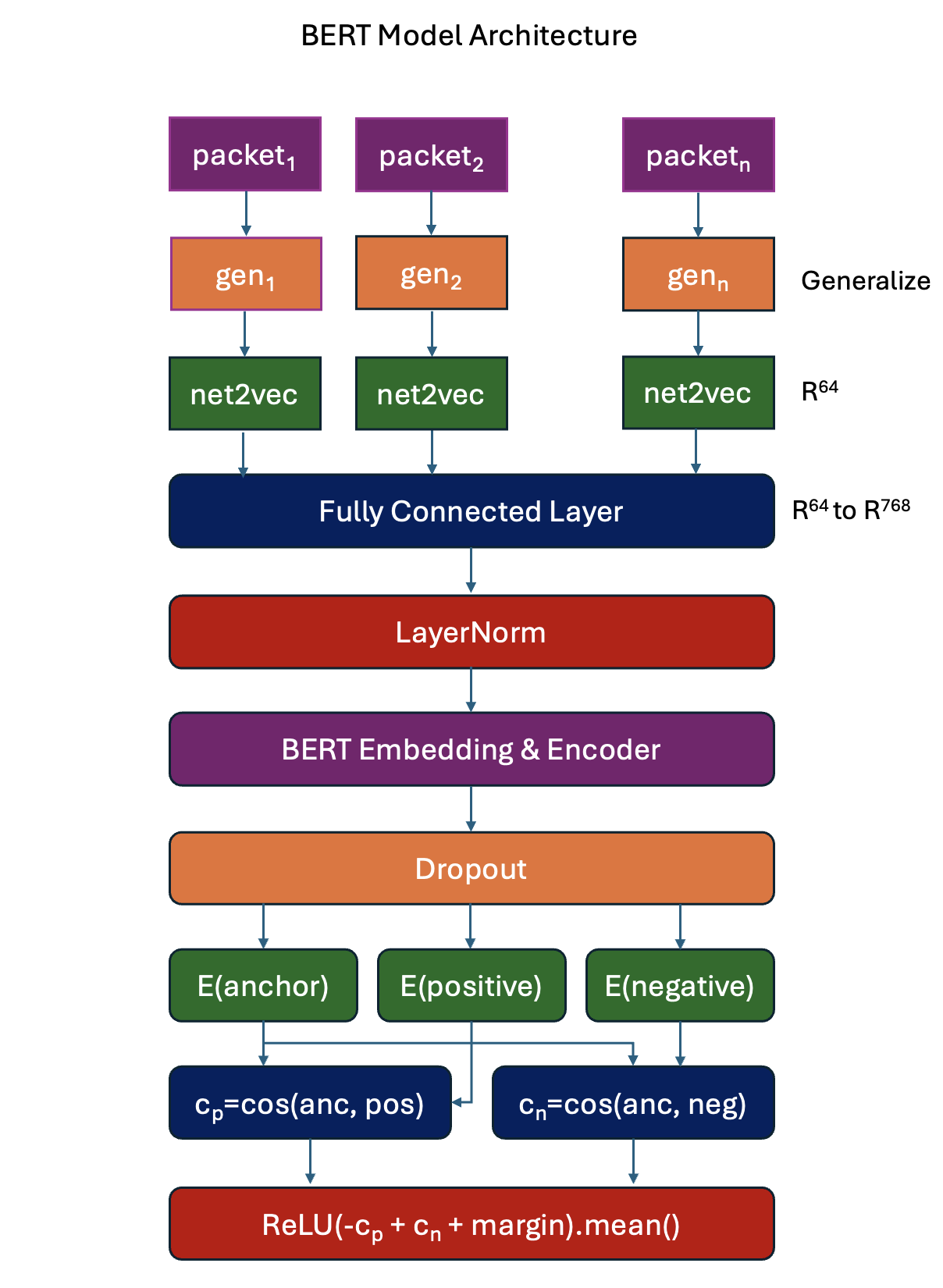}
    \caption{BERT architecture for both models. System calls use sys2vec embeddings; packets use the generalization layer plus net2vec embeddings.}
    \label{fig:bert}
    \end{figure}
    
    Both models use the \texttt{BertModel} class from HuggingFace Transformers~\cite{huggingface}. A fully connected projection layer maps the 64-dimensional sys2vec or net2vec embeddings into the 768-dimensional BERT embedding space, followed by LayerNorm and positional embeddings. After the BERT encoder, the representations are mean-pooled and $\ell_2$-normalized. Cosine similarities between anchors, positives, and negatives are passed into the margin-based contrastive loss from Carter et~al.~\cite{carter2025contrastbert}. The margin is set to 0.1 for both models. The batch size is set to 32 for both models and a learning rate of $1\times 10^{-6}$ is used. Additionally, to help with overfitting, we use $p_{\text{dropout}} = 0.1$ and weight decay of $\lambda = 0.1$.The full architecture is shown in Figure \ref{fig:bert}.
    
    \begin{equation}
    \label{eq:loss}
    \mathcal{L} = \mathbb{E} \left[ \max\left(0,\ -\cos(\mathbf{a}, \mathbf{p}) + \cos(\mathbf{a}, \mathbf{n}) + \text{margin} \right) \right]
    \end{equation}
    
    \subsection{Anomaly Scoring with Isolation Forest}
    Our BERT models are representation learners: during inference they
    produce fixed-dimensional embeddings for each 100-token window, but do not
    directly output anomaly labels. To convert these embeddings into anomaly scores,
    we use an Isolation Forest (IF), a standard unsupervised anomaly detector widely
    used in representation-learning pipelines.
    The IF is trained solely on embeddings from benign data. For the
    system call model, we follow Carter et al.~\cite{carter2025contrastbert} and
    fit the IF using pairwise cosine-similarity features derived from benign
    embeddings. For the packet-based model, we found that using the raw benign
    embeddings yields more stable performance, and therefore train the IF directly
    on the embedding vectors.
    At inference time, each window is embedded and passed through the
    trained IF, which assigns an anomaly score based on the ease with which the sample is
    isolated relative to benign training data. The contamination parameter of the
    IF corresponds to the expected fraction of anomalies (i.e., the target false
    positive rate), allowing us to evaluate performance at multiple operating points.
    We report results at contamination levels (false-positive rates) 0.005, 0.015, and 0.025.
    This anomaly-scoring stage forms the final step of our detection
    pipeline: (1) tokenize system calls or packets, (2) embed using sys2vec or
    net2vec, (3) encode using the contrastively-trained BERT model, and (4)
    score using an Isolation Forest trained on benign data only.
    
    \subsection{Online Anomaly Detection}
    
    In a real deployment, unlike offline analysis, PID information is not assumed and is 
    not used. The router observes a single mixed stream of system-level events and 
    network packets produced by all processes. For this setting, the model operates 
    directly on this interleaved stream: events are embedded, windows are formed 
    sequentially over time, and anomaly scores are produced without any reference 
    to PIDs. The unfiltered mixed-stream online processing for 
    deployment ensures that the offline results measure intrinsic detector quality 
    while the online pipeline reflects the practical operating mode of a real router.To support this deployment scenario, we implement an online anomaly detection 
    pipeline that computes anomaly scores over a live stream of embedded tokens. 
    Raw scores are smoothed using an exponential moving average (EMA), which reduces 
    sensitivity to short-lived benign bursts and stabilizes the decision boundary.

    
    \section{Experimental Setup}\label{sec:ecosystem}
    This section describes the IoTOwl home network ecosystem, the devices and communication protocols involved, the malware behavior patterns executed in the environment, and the data collection procedure used to obtain the system call and packet traces evaluated in this work.

    \subsection{Network Ecosystem}
    
    The IoTOwl testbed is a small home-network ecosystem built around a consumer-grade router configured to serve as the gateway for all connected devices. All system call and packet telemetry used in this study originate from this device, while a lightweight cloud dashboard is used only for visualizing activity.
    The ecosystem includes six heterogeneous IoT and user devices connected through three common home-network protocols (WiFi, Bluetooth Low Energy, and Zigbee). The devices of our testbed include a PurpleAir PA-II air quality sensor (WiFi), a BerryMed pulse oximeter (BLE), a Philips Hue light bulb (Zigbee), a Google Home (WiFi), a smartphone (WiFi), and a laptop (WiFi).
    
    Each of the sensors first connects to the router using its respective protocol, and then the server runs several programs that get the current sensor readings from the devices and pushes them to a Grafana dashboard. The combination of gateway programs that get sensor readings and the AWS Grafana server encompasses an IoT ecosystem called IoTOwl, first introduced by Carter et al. \cite{carter2025contrastbert}

    \subsection{System Call and Network Packet Collection using SkyShark}
    
    \begin{figure}[!b]
    \centering
    \includegraphics[width=0.5\textwidth]{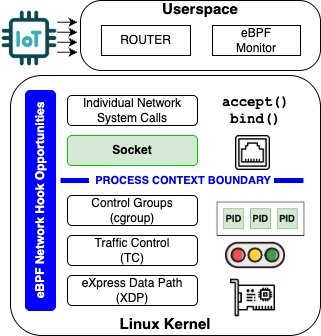}
    \caption{Linux kernel eBPF attachment points for system call and network observability. We intercept data at both the Socket and TC layers to capture complete network context with process attribution.}
    \label{fig:ebpf}
    \end{figure}
    
    To collect high-fidelity system call and network telemetry from the router, we use the Extended Berkeley Packet Filter (eBPF) subsystem~\cite{miano2023fast} in a tool called SkyShark, first developed in the work by Carter et al.~\cite{carter2025contrastbert}. eBPF enables dynamic instrumentation of the Linux kernel and allows real-time monitoring of system call and network events without modifying system binaries.
    
    While Carter et al.~\cite{carter2025contrastbert} focused exclusively on system call collection, we extend SkyShark to gather network packet data as well. Figure~\ref{fig:ebpf} shows the eBPF attachment points we use in the kernel's networking stack. Because attribution of network activity to userspace processes is essential for our anomaly detection approach, we instrument the socket layer to obtain process identifiers (PIDs) for packets generated by userspace applications. To complement the socket-layer view, which does not expose lower-layer protocol fields, we also instrument the Traffic Control (TC) layer to capture full Layer~2--4 header information for all ingress and egress packets.
    
    On its own, the TC layer lacks PID attribution, while the socket layer lacks lower-layer visibility. By correlating information from both layers, our collection pipeline obtains packet traces that include both (i) PID information when available and (ii) complete protocol headers. This dual-layer collection strategy provides the raw system call and packet sequences that are later transformed and modeled for the anomaly detection pipeline.
    
    \subsection{Malware Patterns}
    
    We implement eight malware behaviors following the specifications of prior work~\cite{carter2025contrastbert} to enable direct comparison. Each malware sample runs for five minutes. To capture the malware PIDs for evaluation, we use standard tools such as \texttt{pstree}. The number of observations per malware type is shown in Table~\ref{tab:datasets}.
    We group malware patterns into two categories: (1) OS-focused behaviors that leave no network footprint, and (2) network-focused behaviors that can be detected using both system calls and packet abstractions.
    
    \indent The first OS-focused malware is \textbf{traverse}, which walks the filesystem and issues a \texttt{stat} call for each file, emulating reconnaissance malware seeking sensitive information. The second is \textbf{encrypt}, which similarly traverses the filesystem but encrypts/decrypts each file using \texttt{gpg}, reflecting common ransomware behaviors. \textbf{rename} walks the filesystem and repeatedly renames files, mimicking malware attempting to hide or stage files. \textbf{compile} compiles small C programs and deletes them, representing malware that compiles payloads after landing on a target device.
    
    \indent The first network-focused malware is \textbf{download}, which fetches code from a \texttt{git} repository, representing malware attempting to obtain remote payloads. The next is \textbf{lateral}, which searches for credentials in locations such as \texttt{/etc/shadow} and attempts \texttt{ssh} connections to internal hosts, combining reconnaissance and lateral-movement behavior. \textbf{combo} merges multiple behaviors: downloading code, compiling code, and traversing and encrypting the filesystem, representing a multi-stage infection chain. Finally, the \textbf{APT} is a client/server application that exfiltrates data to an attacker-controlled server while rate-limiting bandwidth and sleeping between bursts to remain stealthy. In our setup, it transmits for 60 seconds and sleeps for 10 seconds with a 100 kB/s rate limit, still producing substantial network activity (Table~\ref{tab:datasets}).
    
    \subsection{Dataset Characteristics}
    
    We evaluate our models using system call and network packet traces collected from 
    the IoTOwl router ecosystem described in Section~4.1. The benign data consist of 
    four separate traces: one 75-minute capture used for training, and three 15-minute 
    captures used exclusively for evaluation. Using multiple benign evaluation sets 
    allows us to measure the generalizability of the detector across independent, 
    separately collected benign periods rather than relying on a single benign baseline.
    
    The primary training set is the \texttt{benign-75min} trace. We truncate the system call portion of this 
    trace to 1{,}000{,}000 calls to reduce training time while preserving behavioral 
    diversity, while all packet abstractions observed during this period are included to 
    build the full net2vec vocabulary (70 tokens total). The truncation is a computational compromise rather than an 
    assumption about sufficiency. The full 13.8M system call trace is highly repetitive due 
    to device polling and steady-state router activity; in practice, the first 
    1M calls contain the range of benign behavior observed across the full trace. 
    Nevertheless, this choice limits the statistical diversity of the training 
    distribution and should be interpreted as a practical constraint rather than a 
    theoretical guarantee. Three additional benign traces (\texttt{benign-15min}, \texttt{benign2-15min}, 
    and \texttt{benign3-15min}) are used for evaluation. These traces were collected 
    at different times of the day under normal router operation, ensuring natural 
    variability in the behavior of the household device. 
    
    Each malware sample is executed for 5 minutes. During evaluation, we retain only 
    the system calls and packets associated with the malware process identifiers 
    (PIDs), which isolates the behavioral footprint of the malware from background 
    traffic. This results in varying observation counts across malware types 
    depending on their activity level (Table~\ref{tab:datasets}). Several malware samples produce small numbers of observations 
    (e.g., a few hundred packets), such as download, which only performs an isolated call to \texttt{git} before exiting. Consequently, extreme detection values 
    (e.g., from 0\% to 100\%) should be interpreted as single-case results rather than 
    as statistically stable estimates. Expanding the number and diversity of malware 
    samples is an important direction for future work.
    
    Table~\ref{tab:datasets} summarizes the total number of system call and packet 
    observations used for training and for evaluation. The separation between 
    training (one long benign trace) and testing (three independent benign traces 
    plus malware traces) ensures that the model is evaluated on behavior it has 
    never seen before, matching the anomaly-detection objective and avoiding 
    any form of leakage across data splits.

    \begin{table}[ht]
    \caption{Number of data observations. Malware counts reflect PID filtering.}
    \label{tab:datasets}
    \centering
    \begin{tabular}{lrr}
    \hline
    \textbf{Dataset} & \textbf{Syscall Count} & \textbf{Packet Count} \\
    \hline
    benign-75min     & 1,000,000 & 1,434,100 \\
    benign-15min     & 2,803,508 &   342,989 \\
    benign2-15min    & 2,663,153 &   228,168 \\
    benign3-15min    & 2,733,216 &   254,723 \\
    \hline
    lateral-5min     &     4,167 &     1,406 \\
    apt6010-5min     &    54,863 &    24,291 \\
    download-5min    &     2,191 &       180 \\
    combo-5min       &    99,952 &     1,221 \\
    rename-5min      &    29,443 &       --- \\
    traverse-5min    &    40,553 &       --- \\
    encrypt-5min     &   147,564 &       --- \\
    compile-5min     &       704 &       --- \\
    \hline
    \end{tabular}
    \end{table}
    
    \subsection{Evaluation Settings and Metrics}
    
    After the model is trained, the evaluation data are passed through the model to calculate transformer embeddings for the evaluation data. We then use an isolation forest to determine how well the model is able to learn benign device behavior. Three benign datasets of 15 minutes each are used in the evaluation of each malware dataset. This is done because we want to gauge how generalizable the approach is across independently collected benign traces and to ensure that the model is not just learning the patterns of a specific benign dataset. Table~\ref{tab:datasets} summarizes the sizes of datasets for both training and testing the model \footnote{\url{https://github.com/anonymous-researcher-520/dimva2026-data}}
.
    
    \indent For the system call–based model, we follow the same evaluation protocol as Carter et al.~\cite{carter2025contrastbert}: pairwise cosine similarities between benign windows are used as input to fit the isolation forest. This preserves comparability to the previous study.
    However, for the packet-based model, we only use the raw benign embeddings from the transformer encoder to fit the isolation forest, which we found produced more stable results for packet sequences.
    
    We use the isolation forest implementation in Scikit-learn~\cite{scikit-learn}, using the default parameters except for the contamination rate and random state for reproducibility. We evaluate detection performance at three contamination rates—0.005, 0.015, and 0.025—which correspond to different operating points on the false-positive spectrum.
    
    In an isolation forest, the contamination rate reflects the expected fraction of anomalies in the data and effectively sets the false-positive budget. Reporting results across multiple contamination levels enables us to illustrate the trade-off between false positives and true positives.
    
    In the router-security context, very low false-positive rates are desirable for always-on deployment, but higher contamination settings help illustrate the upper performance bound of each detector.
    
    \subsection{Limitations}
    
    Although the approach yields strong results, there are 
    several limitations that affect the statistical strength of the findings. 
    First, the evaluation is based on a small set of scripted malware behaviors 
    running for short durations, and several samples generate only limited numbers 
    of system calls or packets. As a result, observed detection rates—especially 
    extreme values such as 0\% or 100\%—should be interpreted as case-study 
    outcomes rather than precise population estimates. We do not perform confidence 
    intervals, hypothesis testing, or multiple runs with different random seeds, 
    which limits the statistical rigor of comparisons. Second, the benign training distribution reflects a single 
    router ecosystem with a small number of devices. While this provides a realistic 
    testbed, it restricts the diversity of benign behavior and may not capture 
    households with significantly different patterns. The truncation of the 
    13.8M-call benign trace to 1M calls further reduces diversity, although it was 
    necessary for computational feasibility. Third, the packet abstraction language simplifies 
    fine-grained header fields into broad categories, improving generalization but 
    discarding subtle information that may be discriminative for advanced threats. 
    Finally, extremely stealthy or long-lived malware may evade both system-call– and 
    packet-based detectors, and EMA smoothing in the online detector introduces a 
    tradeoff between stability and responsiveness. These limitations highlight opportunities for future work in expanding dataset 
    diversity, adding statistical confidence measures, and exploring more expressive 
    packet representations.


    \section{Experimental Results}\label{results}
    
    \subsection{Offline Anomaly Detection Results}
    
    To evaluate our research contributions, we examine three separate, yet related, problems in detail. The first is how the introduction of augmented learning in model training can improve system call-based malware detection. These results are shown in Table \ref{tab:syscall-rates}. The second contribution is the addition of network traffic data to augment the system call data, and these results are summarized in Table \ref{tab:network-rates}. The final contribution is the online anomaly detector that validates the offline work in a real-world setting.
    
    Overall, the experimental results show that augmented learning improves the results for both the system call and network detector consistently. Similarly, using network packets as input to the model drastically increases the detection rate at the lowest false positive rate shown in three out of the four network-focused malware cases. 
    
    Table \ref{tab:syscall-rates} describes our system call-based detection results.  
    We show three contamination rates in the isolation forest, where the contamination parameter specifies the expected proportion of anomalies in the data and, therefore, directly controls the false positive rate. The first column restates the results from Carter et al.~\cite{carter2025contrastbert}. The second column shows our replication using the improved eBPF-based sensor, which offers high-fidelity system-call collection and reduces drop rates under load, leading to better performance even without augmentation. The third column shows the results using augmented learning with a mutation rate of 10\%. This shows a noticeable improvement over non-mutated learning at the 0.005 contamination rate for the lateral and combo malware. The last column is the same as the previous column except that the mutation rate used is 20\%. This yields generally worse results at the 0.005 contamination rate compared to the 10\% mutation rate.

    \begin{table}[t]
    \centering
    \footnotesize
    \renewcommand{\arraystretch}{1.1}
    \setlength{\tabcolsep}{3pt}
    
    \caption{System call detection comparing Carter et al.\ (SOTA) model, our replication using an improved syscall sensor (SOTA'), and augmented-learning models. Mutation rates are listed above each column block; bold indicates the best result for each malware under each contamination level.}
    \label{tab:syscall-rates}
    \resizebox{\textwidth}{!}{%
    \begin{tabular}{lcccc|cccc|cccc}
    \hline
    \multirow{2}{*}{\textbf{Malware}}
     & \multicolumn{4}{c}{\textbf{0.005}}
     & \multicolumn{4}{c}{\textbf{0.015}}
     & \multicolumn{4}{c}{\textbf{0.025}} \\
     & SOTA & SOTA' & 0.1 & 0.2
     & SOTA & SOTA' & 0.1 & 0.2
     & SOTA & SOTA' & 0.1 & 0.2 \\
    \hline
    traverse &\textbf{0.00} &\textbf{0.00} &\textbf{0.00} &\textbf{0.00}
             &0.06 &\textbf{100.00} &\textbf{100.00} &\textbf{100.00}
             &\textbf{100.00} &\textbf{100.00} &\textbf{100.00} &\textbf{100.00} \\
    
    encrypt  &11.64 &\textbf{99.32} &\textbf{99.32} &91.53
             &\textbf{100.00} &\textbf{100.00} &\textbf{100.00} &\textbf{100.00}
             &\textbf{100.00} &\textbf{100.00} &\textbf{100.00} &\textbf{100.00} \\
    
    rename   &48.27 &\textbf{100.00} &\textbf{100.00} &\textbf{100.00}
             &\textbf{100.00} &\textbf{100.00} &\textbf{100.00} &\textbf{100.00}
             &\textbf{100.00} &\textbf{100.00} &\textbf{100.00} &\textbf{100.00} \\
    
    compile  &0.00 &\textbf{100.00} &\textbf{100.00} &\textbf{100.00}
             &\textbf{100.00} &\textbf{100.00} &\textbf{100.00} &\textbf{100.00}
             &\textbf{100.00} &\textbf{100.00} &\textbf{100.00} &\textbf{100.00} \\
    
    download &0.00 &50.00 &50.00 &\textbf{75.00}
             &\textbf{100.00} &\textbf{100.00} &\textbf{100.00} &75.00
             &\textbf{100.00} &\textbf{100.00} &\textbf{100.00} &\textbf{100.00} \\
    
    lateral  &33.05 &25.00 &\textbf{100.00} &25.00
             &99.44 &25.00 &\textbf{100.00} &\textbf{100.00}
             &\textbf{100.00} &37.50 &\textbf{100.00} &\textbf{100.00} \\
    
    combo    &0.00 &13.57 &\textbf{56.78} &21.11
             &1.79 &97.99 &\textbf{98.49} &93.47
             &\textbf{100.00} &98.49 &98.99 &98.49 \\
    
    apt6010  &\textbf{0.00} &\textbf{0.00} &\textbf{0.00} &\textbf{0.00}
             &0.00 &0.92 &\textbf{100.00} &0.00
             &\textbf{100.00} &0.92 &\textbf{100.00} &\textbf{100.00} \\
    \hline
    \end{tabular}
    }
    \end{table}

    Table \ref{tab:network-rates} shows results for the four network-focused malware. As with system calls, we compare no mutation, 10\% mutation, and 20\% mutation against the best system call result from Table \ref{tab:syscall-rates}. Overall, the 10\% mutation model performed the best, similar to the system call results, and showed drastic improvements for the download, combo, and APT malware at the 0.005 contamination rate. In one case, the detection even increased from 0\% to 100\%, a remarkable improvement.
    
    \begin{table}[ht]
    \centering
    \footnotesize
    \renewcommand{\arraystretch}{1.1}
    \setlength{\tabcolsep}{3pt}
    
    \caption{Detection rates (\%) for system call–based and network-based
    detections for network-focused malware, where the column headers denote the mutation rate. 
    System call values correspond to Mut=0.1 in Table~\ref{tab:syscall-rates}.
    Best value per malware per contamination rate is in bold.}
    \label{tab:network-rates}
    
    \resizebox{\textwidth}{!}{%
    \begin{tabular}{lcccc|cccc|cccc}
    \hline
    \multirow{2}{*}{\textbf{Malware}} &
    \multicolumn{4}{c}{\textbf{0.005}} &
    \multicolumn{4}{c}{\textbf{0.015}} &
    \multicolumn{4}{c}{\textbf{0.025}} \\
    & Sys & 0.0 & 0.1 & 0.2 
    & Sys & 0.0 & 0.1 & 0.2
    & Sys & 0.0 & 0.1 & 0.2 \\
    \hline
    download &50.00 &0.00 &\textbf{100.00} &0.00 
            &\textbf{100.00} &\textbf{100.00} &\textbf{100.00} &\textbf{100.00}
            &\textbf{100.00} &\textbf{100.00} &\textbf{100.00} &\textbf{100.00} \\
    lateral  &\textbf{100.00} &50.00 &\textbf{100.00} &21.43
            &\textbf{100.00} &\textbf{100.00} &\textbf{100.00} &92.86
            &\textbf{100.00} &\textbf{100.00} &\textbf{100.00} &\textbf{100.00} \\
    combo    &56.78 &0.00 &\textbf{100.00} &0.00
            &98.49 &\textbf{100.00} &\textbf{100.00} &\textbf{100.00}
            &98.99 &\textbf{100.00} &\textbf{100.00} &\textbf{100.00} \\
    apt6010  &0.00 &0.00 &\textbf{100.00} &0.00
            &\textbf{100.00} &0.00 &\textbf{100.00} &0.00
            &\textbf{100.00} &\textbf{100.00} &\textbf{100.00} &0.00 \\
    \hline
    \end{tabular}
    }
    \end{table}
    
    \subsection{Online Anomaly Detection Results}
    
    In the offline evaluation setting, system calls and packets can be filtered by process identifier (PID), enabling accurate anomaly scoring. 
    However, a real deployment on a home router cannot rely on PID filtering, since packet streams and system activity must be analyzed continuously without malware behavior isolation.
    To support this setting, we implement an online anomaly detection pipeline that computes anomaly scores over a live stream of embedded tokens.
    
    Raw scores from the Isolation Forest are smoothed using an exponential moving average (EMA), which reduces sensitivity to short-lived bursts of benign activity and stabilizes the decision boundary in the absence of PID separation. 
    The EMA is defined in Equation~\ref{ema}, where $c_t$ is the current score and $v_{t-1}$ is the previous smoothed value:
    
    \begin{equation}
    v_t = \alpha \cdot c_t + (1 - \alpha) \cdot v_{t-1}.
    \label{ema}
    \end{equation}
    
    The EMA attenuates brief benign spikes—common in IoT polling and background router activity—while still reacting to persistent deviations indicative of malware.
    To avoid manually tuning $\alpha$, we fit a shallow model on benign/malicious samples that predicts an appropriate smoothing weight, ensuring stable online behavior without hand-designed parameters. A final anomaly decision is made by comparing the smoothed score to a threshold set as the $(100 \times \text{contamination})$\textsuperscript{th} percentile of benign training scores, providing a direct mapping between a chosen false-positive budget and the operational boundary. This online evaluation complements the PID-filtered offline results by demonstrating performance in a realistic deployment scenario where mixed benign and malicious activity must be processed continuously.

    \begin{table}[ht]
    \centering
    \footnotesize
    \renewcommand{\arraystretch}{1.1}
    \setlength{\tabcolsep}{3pt}
    
    \caption{TTD results in seconds for lowest FPRs with the number of runs the malware was detected out of the ten experiments in brackets. FPR over 30 minutes for a given contamination rate is provided in the last row.}
    \label{tab:results}
    
    \begin{tabular}{lcc}
    \hline
    \textbf{Malware} & \textbf{0.005} & \textbf{0.015} \\
    \hline
    Traverse & $8.405 \pm 6.892$ [10/10] & $0.049 \pm 0.055$ [10/10] \\
    Encrypt  & $6.347 \pm 3.984$ [10/10] & $0.121 \pm 0.079$ [10/10] \\
    Rename   & $0.521 \pm 0.090$ [10/10] & $0.137 \pm 0.074$ [10/10] \\
    Compile  & $0.470 \pm 0.247$ [10/10] & $0.039 \pm 0.055$ [10/10] \\
    Download & $0.600 \pm 0.523$ [3/10]  & $0.052 \pm 0.051$ [10/10] \\
    Lateral  & $0.191 \pm 0.060$ [10/10] & $0.124 \pm 0.171$ [10/10] \\
    Combo    & $1.397 \pm 1.643$ [6/10]  & $0.108 \pm 0.192$ [10/10] \\
    APT      & --                       & $0.061 \pm 0.040$ [10/10] \\
    \hline
    FPR      & 0.0011                   & 0.0167 \\
    \hline
    \end{tabular}
    \end{table}

    The main metric used to evaluate online anomaly detection is time-to-detection (TTD). The TTD and the false positive rate are the most important statistics to analyze in online anomaly detection because anomaly detection must be both quick and correct. Our results in Table \ref{tab:results} show that all malware patterns can be detected in less than a second after initial malware infection at the 0.015 false positive rate, and half can be detected in less than a second at a false positive rate of 0.005, though not in all ten trials in every case. For the others, three can be detected in a matter of seconds, while the APT cannot be detected. We show the results at both of these false positive rates in order to illustrate the tradeoff between low false positive rate and quicker detection.
    
    \section{Future Work}
    An area to explore in future work is incorporating a BLE sensor and its data.  Network packet data are, in essence, subsets of system call data, since network traffic is seen through system calls such as \texttt{socket}. However, the results in this research show that while that may theoretically be the case, in practice there is too much noise to pinpoint those actions as clearly with system calls as is possible using network packet data. Another area to explore is to try more recent transformer models.  BERT remains a remarkable all-purpose model that is relatively accessible for researchers without the resources of large AI companies, but the vanilla BERT model came out several years ago, which is a long time in today's accelerated AI research timeline. In addition, though the vanilla BERT model yields excellent results, we would like to implement even stealthier malware and more complex ecosystem configurations to test the limits of the model. To this end, alternative models to try are XLNet \cite{yang2020xlnetgeneralizedautoregressivepretraining} and ModernBERT \cite{warner2024smarterbetterfasterlonger}.
    
    \section{Summary and Conclusions}
    This work focuses on advancing the state of the art in behavioral anomaly detection for edge devices. We introduced augmented learning during the training process to show how this improves system call-based malware detection. We then applied a similar pipeline to network traffic packet abstractions, which yields better detection for network-intensive malware at a low false positive rate. We also provided a production-ready online anomaly detection framework to validate our approach.
    
    The system call and network packet data are collected from a router that services a realistic home network consisting of several devices using a variety of communication protocols. Once the packets are collected from the router, they are generalized to packet abstractions. These abstractions are then embedded into 64-length embeddings by a Word2vec-like model we call net2vec. We also tried using fixed random embeddings as part of an ablation study to determine how useful net2vec is, but this yielded worse results. After creating vector embeddings for each packet word in the observed vocabulary, we process the data into sentences of length 500 for syscalls and 100 for network packets, which are then split into triplets. Each triplet contains an anchor, a positive example (similar to the anchor), and a negative example (designed to be dissimilar to the anchor). The random negative is then mutated by changing a portion of the packets in the negative sentence to different packet configurations seen in the training data. 
    Using these triplets and the custom triplet loss function, the BERT model learns to keep the anchor and positive similarity high while attempting to reduce the similarity between the anchor and the negative example.
    The BERT model is then evaluated by fitting an isolation forest on three separate benign datasets and detecting outliers in several types of malware patterns. 
    
    \indent The experimental results show that using augmented learning as part of the training process boosts system call-based detection at low false positive rates. Similarly, a model trained on network data works even better for network-focused malware at low false positive rates, which is expected given the more pronounced footprint of network-focused malware in the network traffic. 
    
    \indent There are three main contributions of this research. The first is the introduction of augmented learning to the BERT training, which builds on the SOTA using only contrastive learning and yields on average a 52 percentage point improvement over SOTA at the 0.005 false positive rate using a 0.1 mutation rate. The second is the introduction of a new network packet abstraction language that provides a similar pipeline for network packets as that of system calls. We show that using a 0.1 mutation rate, we were able to achieve a 48 percentage point improvement on average over our own system call-based detector using a 0.1 mutation rate. Finally, the online detection study shows that the methodology in this work can be applied to a real-world scenario effectively and validates the results obtained in Section \ref{results}. All of these results advance the SOTA.
    
    \section*{Ethical Considerations}
    The goal of this work is to improve security on edge devices. The data was collected in a closed ecosystem using only malware that does not propagate or infect any unintended nodes by design. No malicious code used in this work is shared publicly and no personally identifiable data are used.

    
    \bibliographystyle{splncs04}
    \bibliography{bib}
    
    \end{document}